\documentclass[letterpaper,12pt]{article}   

\usepackage{graphics, epsfig}
\begin{document}

\title{A simple inverter for polarization transformations}


\author{\bf Rajendra Bhandari}
\maketitle
\vspace{10mm}
\begin{center}
\begin{tabular}{ll}
            & Raman Research Institute, \\
            & Bangalore 560 080, India. \\
            & email: bhandari@rri.res.in\\
\end{tabular}
\end{center}
\vspace{10mm}

\begin{center}
\Large{\bf {Abstract}}
\end{center}
 It is shown that a transparent birefringent element whose eigenstates are a pair of elliptical polarizations with principal
axes along the $x$ and $y$ axes, sandwiched between a pair of orthogonal halfwave plates with
principal axes along directions at 45 deg to  the $x$ and $y$ axes, is equivalent to an 
element with the same eigenstates and  eigenvalues  but with the 
fast and the slow eigenstates interchanged. The device thus produces the inverse of the 
original unitary transformation. A similar result holds for a pure
dichroic element. With electrically switchable halfwave plates such a device can be used to 
switch the sign of optical activity or to  rotate through 90 deg, without any moving 
parts, a linear retarder or a linear polarizer.
\vspace{5mm}

OCIS Codes 260.5430, 230.4110, 230.2090, 280.4788
\newpage
 

 In this paper we  point out an interesting application in polarization optics of some 
well known results in the theory of 2x2 complex matrices. It is  well known that any 
non-depolarizing polarization transformation can be represented by a 2x2 complex matrix (Jones matrix) 
and that any 
such matrix $S$ can be expressed as a product $S=UD$ or $S=D'U$, where $U$ is a unitary matrix 
and $D, D'$ are nonunitary matrices with orthogonal eigenstates and real eigenvalues \cite{grouptheory,luchapman}. 
We shall choose 
the first of the two decompositions and $D$ will be called 
a ``pure dichroic element". The matrix $U$ can be written as a product of an isotropic phase 
factor exp($i\alpha$) and a matrix with  determinant $+1$ whose eigenvalues are exp(-$i\delta$) and 
exp($i\delta$). Similarly the matrix $D$ can be written as a product of an isotropic attenuation
factor exp(-$\eta$) and a matrix with unit determinant with eigenvalues exp(-$\gamma$) and 
exp($\gamma$). From now on we shall ignore the isotropic factors and assume that $U$ and $D$ 
are matrices with unit determinant, i.e. elements of the group SL(2,C); the former also being an 
element of the SU(2) subgroup of SL(2,C). We note that for an arbitrary element of SL(2,C) 
the sets of orthogonal eigenstates of $U$ and $D$ are different.

Let us first consider a unitary transformation $U$. Let $|u>$ and $|\tilde u>$ be 
the two orthogonal eigenvectors of $U$ with eigenvalues exp$(-i\delta)$ and exp$(i\delta)$ 
respectively. Let $A$ be some other element of SU(2) such that $A|u>=|u'>$ and $A|\tilde u>=|\tilde u'>$.
It is then easy to verify that the matrix $U'=AUA^\dagger$ has the states $|u'>$ and $|\tilde u'>$ as its 
eigenvectors, with eigenvalues exp$(-i\delta)$ and exp$(i\delta)$ respectively. The matrix $A^\dagger$ is 
the hermitian conjugate of $A$, which is the same as $A^{-1}$ when $A$ is unitary. Now if the matrix $A$ 
is chosen such that $|u'>=|\tilde u>$ and $|\tilde u'>=|u>$ then we have the result that the matrix 
$AUA^\dagger$ has the states $|\tilde u>$ and $|u>$ as its eigenstates with eigenvalues 
exp$(-i\delta)$ and exp$(i\delta)$ respectively.  $U'$ therefore represents the 
transformation inverse of that represented by $U$. For example  if  $U$ 
corresponds to pure optical activity then the matrix $U'=AUA^\dagger$ also corresponds to pure optical activity 
of the same magnitude but with the sign changed. If on the other hand $U$ represents a linear retarder with its fast axis 
oriented at an angle  $\psi$, 
the matrix $AUA^\dagger$ represents a linear retarder with the same retardation but with its fast axis oriented at $\psi + 90^\circ$. 

It is not difficult to see what the matrix $A$ is. A state $|u>$ represented by the point $(\theta,\phi)$ 
on the Poincar\'{e} sphere can be taken to its orthogonal state $|\tilde u>$ by means of a $\pi$ rotation about the 
point $(90^\circ,\phi+90^\circ)$. This corresponds to a halfwave plate whose fast axis makes, in real 
space, an 
angle $45^\circ$ with the major axis of the polarization ellipse corresponding to the state $(\theta,\phi)$  on the 
Poincar\'{e} sphere. This immediately leads to the first main result of the paper, i.e. a unitary 
polarization element $U$ corresponding to a rotation through $2\delta$ about a point $(\theta,\phi)$ on 
the Poincar\'{e} sphere, sandwiched between two orthogonal halfwave plates whose fast axes 
make, in real space, angles $\pm 45^\circ$ with the major axis of one of the ellipses representing the eigenpolarizations 
of $U$ will correspond to a rotation through an angle $-2\delta$ about the point $(\theta,\phi)$ on the sphere . 
By orthogonal halfwave plates we mean those whose fast axes make an angle $90^\circ$ with 
each other.

Let us next consider a dichroic transformation represented by a pure dichroic element $D$, i.e. 
an element with orthogonal polarization eigenstates  $|d>$ and $|\tilde d>$ and eigenvalues 
exp$(-{\gamma})$ and exp$({\gamma})$ respectively. It can be shown exactly as above that such 
an element, sandwiched between two orthogonal halfwave plates whose fast axes 
make angles $\pm 45^\circ$ with the major axis of one of the ellipses representing the eigenpolarizations 
of $D$ will correspond to a transformation which is inverse of $D$, i.e. have $|\tilde d>$ and $|d>$ 
as eigenstates with eigenvalues exp$(-\gamma)$ and exp$(\gamma)$ respectively; $\gamma$ 
being a real quantity. If $D$ corresponds to a linear polarizer, the sandwich corresponds to the same 
linear polarizer, rotated through $90^\circ$.  

With the use of electro-optic modulators, i.e elements which can be switched electrically from being unit elements to being 
halfwave plates, the above considerations lead to the possibility of remotely inverting a 
unitary or a pure dichroic polarization transformation, for example  (i)  switching 
the sign of optical activity of an optically active sample, (ii)  rotating 
the principal axes of a linear retarder through $90^\circ$ or (iii)  rotating through $90^\circ$ 
the transmission axis of a linear polarizer, all the operations being without the use of any moving parts.

It is easy to show that the above device inverts not only unitary and pure dichroic transformations 
but also inverts a product $S=UD$ of the two kinds of transformation if $S$ and $D$ have the same orthogonal pair 
of eigenstates. For then, since $U$ and $D$ commute,

\begin{eqnarray}
  S'=ASA^\dagger=(AUA^\dagger)(ADA^\dagger)=U^{-1}D^{-1}\nonumber \\
=(DU)^{-1}=(UD)^{-1} =(S)^{-1} \label{eq:c1}
\end{eqnarray}

\noindent The above equation implies that if a linear polarizer has a small amount 
of linear birefringence with the same principal axes as those of the polarizer, the 
device described above would still work, i.e. the  imperfect 
polarizer, sandwiched between a pair of orthogonal halfwave plates at $\pm 45^\circ$ is 
equivalent to the same imperfect polarizer, rotated through $90^\circ$.

While  the result described in this paper is trivial as a mathematical result
we have been unable to find in literature an instance of its use in 
polarization optics, hence this communication.



\end{document}